\documentclass[11pt,dvips]{article}

\usepackage{epsfig,times} 
\usepackage{picinpar}
\usepackage{floatflt}
\makeatletter
\renewcommand{\section}{\@startsection{section}{1}{0in}
        {0.4\baselineskip}{0.1\baselineskip}{\Large\bf}}
\renewcommand{\subsection}{\@startsection{subsection}{2}{0in}
        {0.25\baselineskip}{-\baselineskip}{\large\bf}}
\renewcommand{\subsubsection}{\@startsection{subsubsection}{3}{0in}
        {0.1\baselineskip}{-\baselineskip}{\normalsize\bf}}
\makeatother
\begin{document}

%
\thispagestyle{myheadings}
%
\markright{OG 2.1.18}
\begin{center}
%
{\LARGE \bf Evidence for a QPO structure in the TeV and X-ray light curve
  during the 1997 high state $\gamma$ emission of Mkn 501}
\end{center}

\begin{center}
%
%
{\bf D. Kranich$^1$, O.C. De Jager$^2$, M. Kestel$^1$, E. Lorenz$^1$,
  and the HEGRA collaboration}\\
{\it $^1$ Max-Planck-Institut f\"ur Physik, M\"unchen, D-80805,
  Germany\\
  $^2$ Potchefstroom University for CHE, Potchefstroom 2520, South Africa}
\end{center}

\begin{center}
{\large \bf Abstract\\}
\end{center}
\vspace{-0.5ex}
%
%
The BL Lac Object Mkn 501 was in a state of high activity in the TeV
range in 1997. During this time Mkn 501 was observed by all
Cherenkov-Telescopes of the HEGRA-Collaboration. Part of the data were
also taken during moonshine thus providing a nearly continuous
coverage for this object in the TeV-range. We have carried out a
QPO analysis and found evidence for a 23 day periodicity.
We applied the same analysis on the 'data by dwell' x-ray lightcurve
from the RXTE/ASM database and found also evidence for the 23 day
periodicity. The combined probability was ${\cal P} = 2.8\, e$--$\, 04$.
%

\vspace{1ex}

\section{Introduction}
The nearby AGN Mkn 501 ($z = 0.034$) of the BL-Lac class is known to
be one of the few TeV $\gamma$-ray (shortcut $\gamma$) sources. 
The source has been discovered in 1995 by the Whipple collaboration,
Quinn et al. 1996, at a flux level equivalent to a few \% of
that of the Crab nebula and has been confirmed 1996 by the HEGRA
collaboration, Bradbury et al. 1997. 

Early in 1997 a large increase in the TeV $\gamma$-flux has been
observed by at least 7 groups, see Protheroe 1997 for an
overview. Observations covered the time from February until
October. The flux was highly variable with a mean value of around 3
times and flares up to 10 times of that of the Crab.

Since 1996 Mkn 501 has been monitored in the 2-12 (10) keV X-ray band
by the all sky monitor (ASM) of RXTE (Remillard 1997). The
data are available as public domain (RXTE 1999) and
presented either on rates per few hours or smoothed. The Mkn 501 X-ray
data coverage is nearly continuous (with small gaps) and shows also
a significant increase until mid 1997 and then a falling tendency.
Again, significant fluctuations are observed, but much less dramatic
compared to the TeV flux variation.

Already at the XXV ICRC, Durban, two groups (TA and HEGRA) reported
about indications for periodicity in the TeV lightcurve, although no
full analysis has been carried out. One of the problems was the
interruption of the observations during the periods of moonshine. In
the following the procedure and results of a QPO analysis to both the
TeV and X-ray lightcurve will be presented.

\section{The data}
The TeV data were recorded with the 6 air Cherenkov telescopes of the
HEGRA experiment located on the canary island La Palma. Four 
telescopes formed a stereo system (SCT) providing high statistics
results from 110 hours of observation time. Less precise data were
taken with the two standalone CTs. The latter instrument's data
correspond to 300 hours observation, in part overlapping with the SCT
data. A sizable fraction of the data recorded with one of the
standalone telescopes, CT1, was taken during moonshine, thus
providing flux measurements when no other telescope was operational,
see Aharonian et al. 1999a, Aharonian et al. 1999b, Kranich et al. 1999. 

\begin{figure}
  \epsfig{file=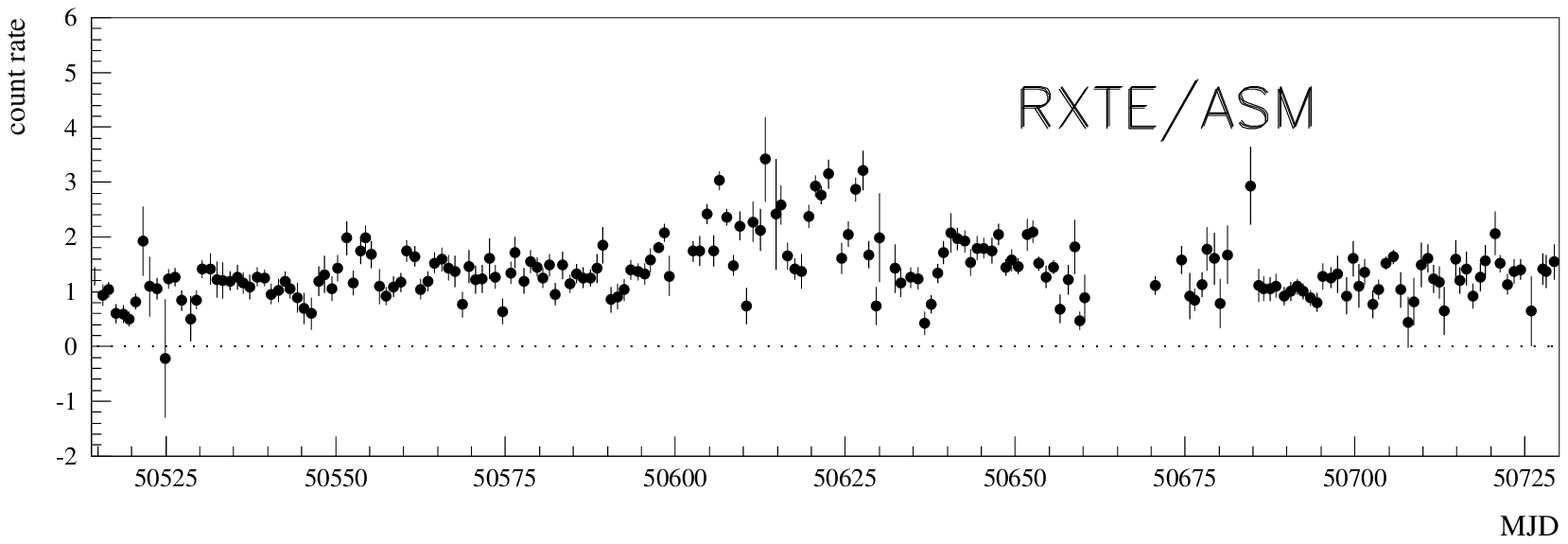,height=5.cm,width=\linewidth}
  \epsfig{file=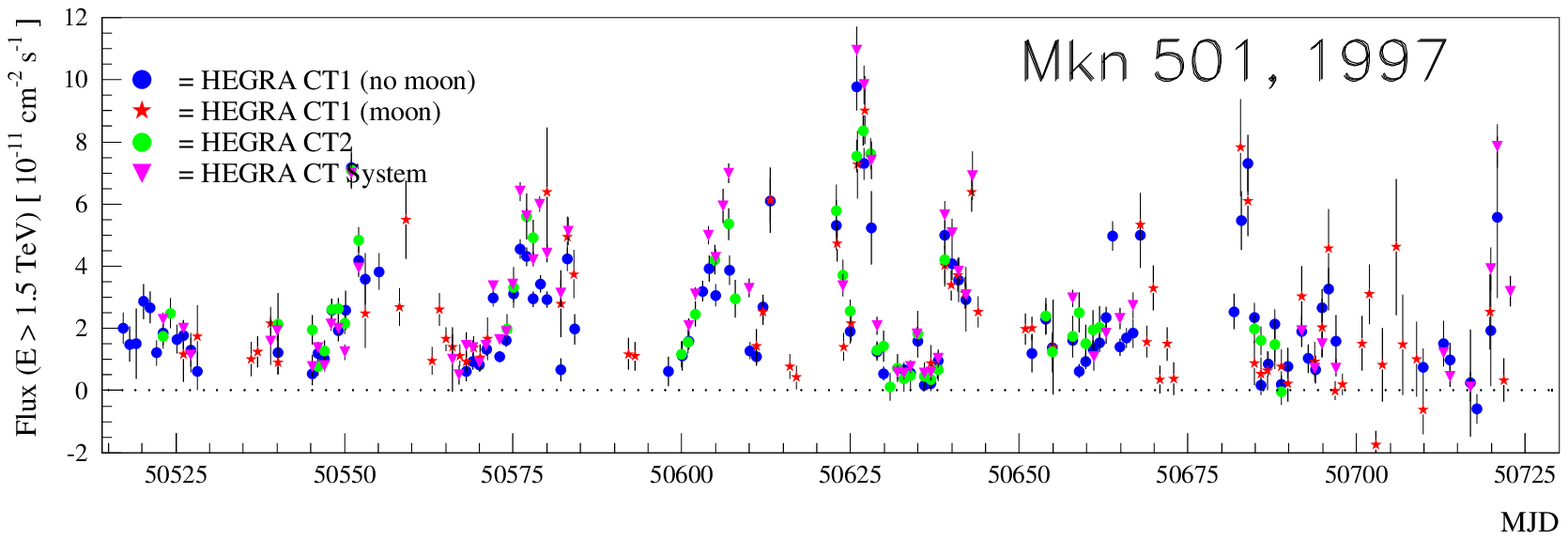,height=5.cm,width=\linewidth}
  \caption{\label{usedlightcurve}
    Lightcurve above 1.5 TeV energy; also shown is the X-ray
    lightcurve between 2 to 10 keV from RXTE-ASM. The one day
    averaged RXTE data are shown for clarity.}
\end{figure}

The used lightcurve above 1.5 TeV energy is shown in
Fig.~\ref{usedlightcurve}. Typical observation duration during summer
1997 was 3-4 hours/night whenever atmospheric conditions were
acceptable. Fig.~\ref{usedlightcurve} also shows the X-ray lightcurve
between 2 to 10 keV from RXTE-ASM. For clarity the smoothed data are
shown while in the subsequent analysis the unsmoothed data ($\sim
3000$ data points) were used.

For the analysis the data between MJD 50545 and MJD 50661 were used. 
As starting date we took the onset of the high state while the second
limit was taken where the X-ray curve was interrupted for some
time. Additional reason to end the analysis at MJD 50661 was the fact
that the later TeV data were taken only at large zenith angles and
rather short observation times per day. Also the source seemed to
return to a lower state.

\section{The periodicity analysis}

As mentioned in the introduction and visible in Fig.~\ref{usedlightcurve} 
the TeV lightcurve shows indications of a periodic modulation. In
order to test for a QPO structure we used the formalism developed by
Lomb and Scargle (Lomb 1976, Scargle 1982) which is able to derive power
spectra from unevenly sampled data. The Lomb {\it normalized periodogram}
(nP) is a modification of the classical Periodogram (defined within the
formalism of discrete Fourier transformations) and gives the spectral power
as a function of angular frequency $\omega \equiv 2 \pi f$. The nP is
defined as:

\begin{equation}
P_N \left( \, \omega \, \right) = {1 \over 2 \, \sigma^2}
\left\lgroup { \left[ \sum_j (h_j - \overline{h}) \cos\, \omega \left(t_j - \tau \right)
\right]^2 \over \sum_j \cos^2\, \omega \left(t_j - \tau \right)} +
{ \left[ \sum_j (h_j - \overline{h}) \sin\, \omega \left(t_j - \tau \right)
\right]^2 \over \sum_j \sin^2\, \omega \left(t_j - \tau \right)}
\right\rgroup\
\end{equation}
with $\tau$ defined by the equation:
\begin{equation}
\tan\,\left(2\,\omega\,\tau\,\right) = \Bigl(\,\sum_j \sin\,2\,
\omega\,t_j\,\Bigr) \ / \ \Bigl(\,\sum_j \cos\,2\,\omega\,t_j\,\Bigr)
\end{equation}
The $h_j$ are the individual flux values measured at time $t_j$, and
the sums extend over all $N$ data points. $\overline{h}$ is the mean
value of the $h_j$.\\
This definition of $\tau$ makes $P_N \left( \, \omega \, \right)$
completely independent of a constant time shift of the data.
It can be shown (Scargle 1982) that the normalization on
the variance forces $P_N \left( \, \omega \, \right)$ to have an
exponential probability distribution if the data values are
independent Gaussian random values:
\begin{equation}
{\cal P} \left(\, P_N \left( \, \omega \, \right) > z \, \right)
\ \equiv \ \exp \,(-z) \ .
\label{prob_equ}
\end{equation}

\begin{figwindow}[1,r,%
{\mbox{\epsfig{file=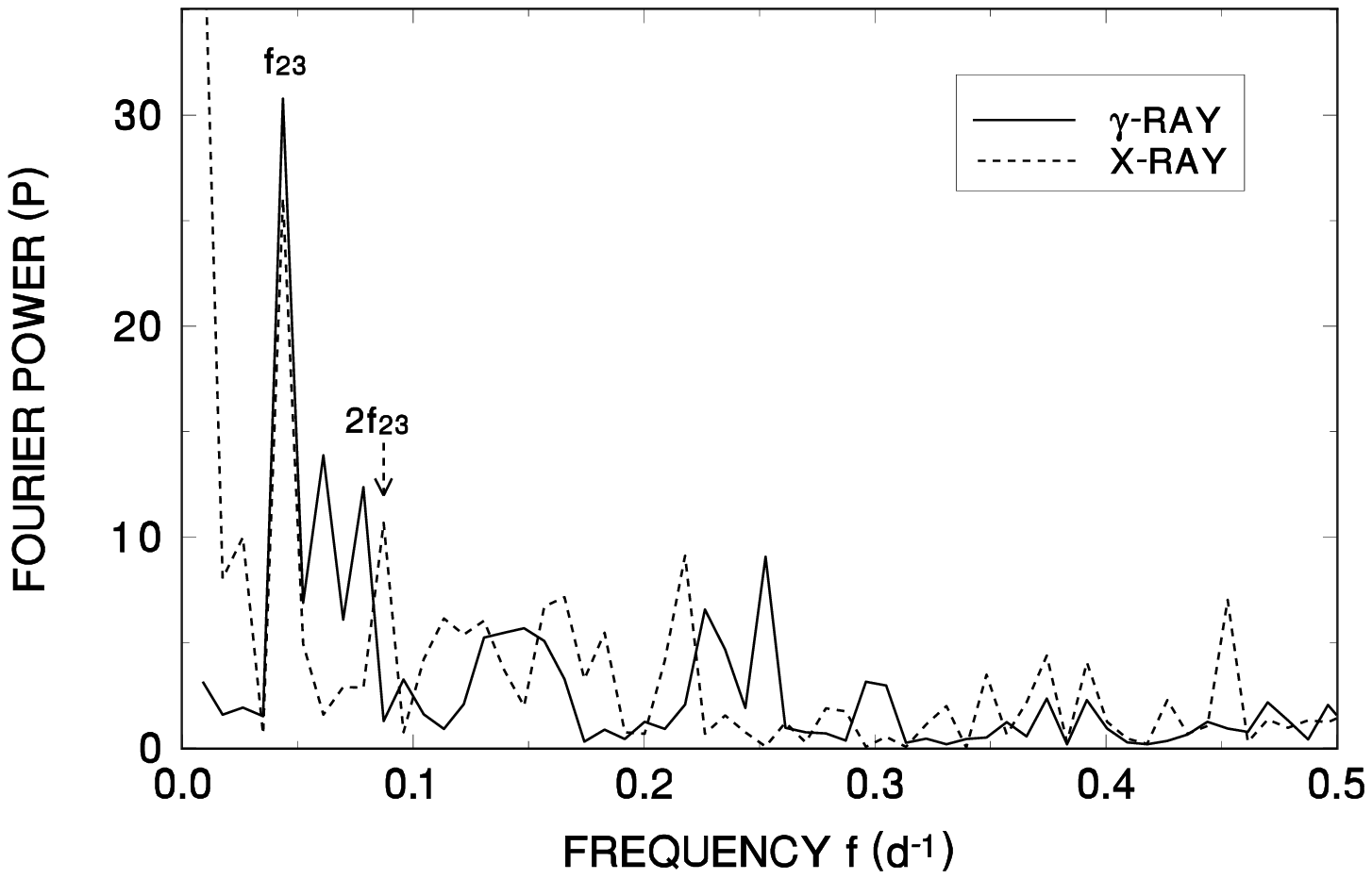,width=10.0cm,height=6.5cm}}},%
{\label{power_spectra}
Power spectra for TeV (solid line) and x-ray data (dashed line).}]
The results from the analysis, carried out independently for the TeV
and X-ray data set, are shown in Fig.~\ref{power_spectra}.
Both, the TeV and X-ray data show a significant peak at a frequency
of $0.044 \, d^{-1}$ which corresponds to a 23 day period. The 2nd
harmonic of this period is visible in the X-ray data but not in the TeV
data. Note that the high power value in the X-ray data at $f = 1 \,
/ \, T = 0.0088\, d^{-1}$ corresponds to the overall behavior of the
average count-rate during 1997 (increasing/decreasing count rates
before/after  MJD $\sim$ 50625).\\
From the power value for the 23 days period one can calculate
the corresponding probability against the null hypothesis of
independent Gaussian random noise. But since it is known that BL Lac
objects show the phenomenon of flaring it is more meaningful to test
against the null hypothesis of independent, random distributed flares
(so-called shot noise model, see deJager 1999 and reference therein).
In order to state probabilities within this framework it is necessary
to renormalize nP. This can be done by calculating the mean Fourier
power from the shot noise model. If one assumes flares of Gaussian shape
with standard deviation $\sigma$ the mean Fourier power
becomes: 
\end{figwindow}

\begin{equation}
\left< \, P_f \left( \, \omega \, \right)\, \right> = A \cdot \exp \,
\left( \, - \omega^2\, \sigma^2 \right) 
\end{equation}
Here $A \, \propto \, \overline{a^2}$ which is the mean squared amplitude
of the flares.
A maximum likelihood analysis of the derived power spectra of
Fig.~\ref{power_spectra}, but excluding the 23 day period and the first
frequency $f = 1 \, / \, T$, gives the parameters $A = 3.8$, $\sigma = 0.41\,
{\rm d}$ (TeV range) and $A = 3.8$, $\sigma = 0.35\, {\rm d}$ (X-ray range)
(deJager 1999). The final Fourier power is then calculated as:
\begin{equation}
P \left( \, \omega \, \right) =
P_N \left( \,  \omega \, \right)  / \left< \, P_f \left( \, \omega \,
  \right)\, \right> \, .
\end{equation}

\begin{figure}
\begin{minipage}[b]{.48\linewidth}
\centering\epsfig{file=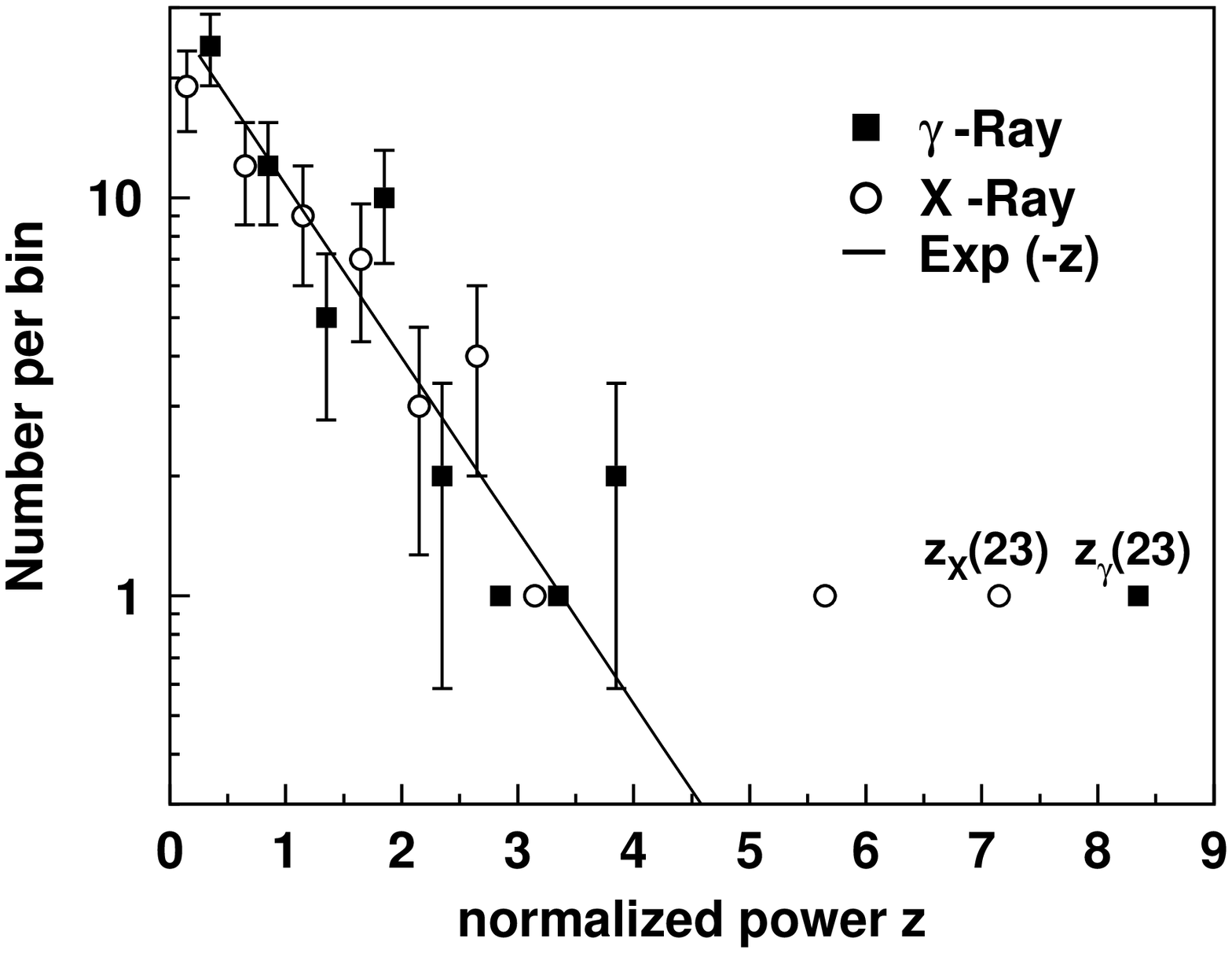,width=\linewidth,height=6.cm}
\caption{\label{norm_power_spectra}
Number of frequencies with a given (renormalized) power vs. power for
the TeV (black squares) and X-ray data (open circles). Points without
errors consist of one single frequency.}
\end{minipage}\hfill
\begin{minipage}[b]{.48\linewidth}
\centering\epsfig{file=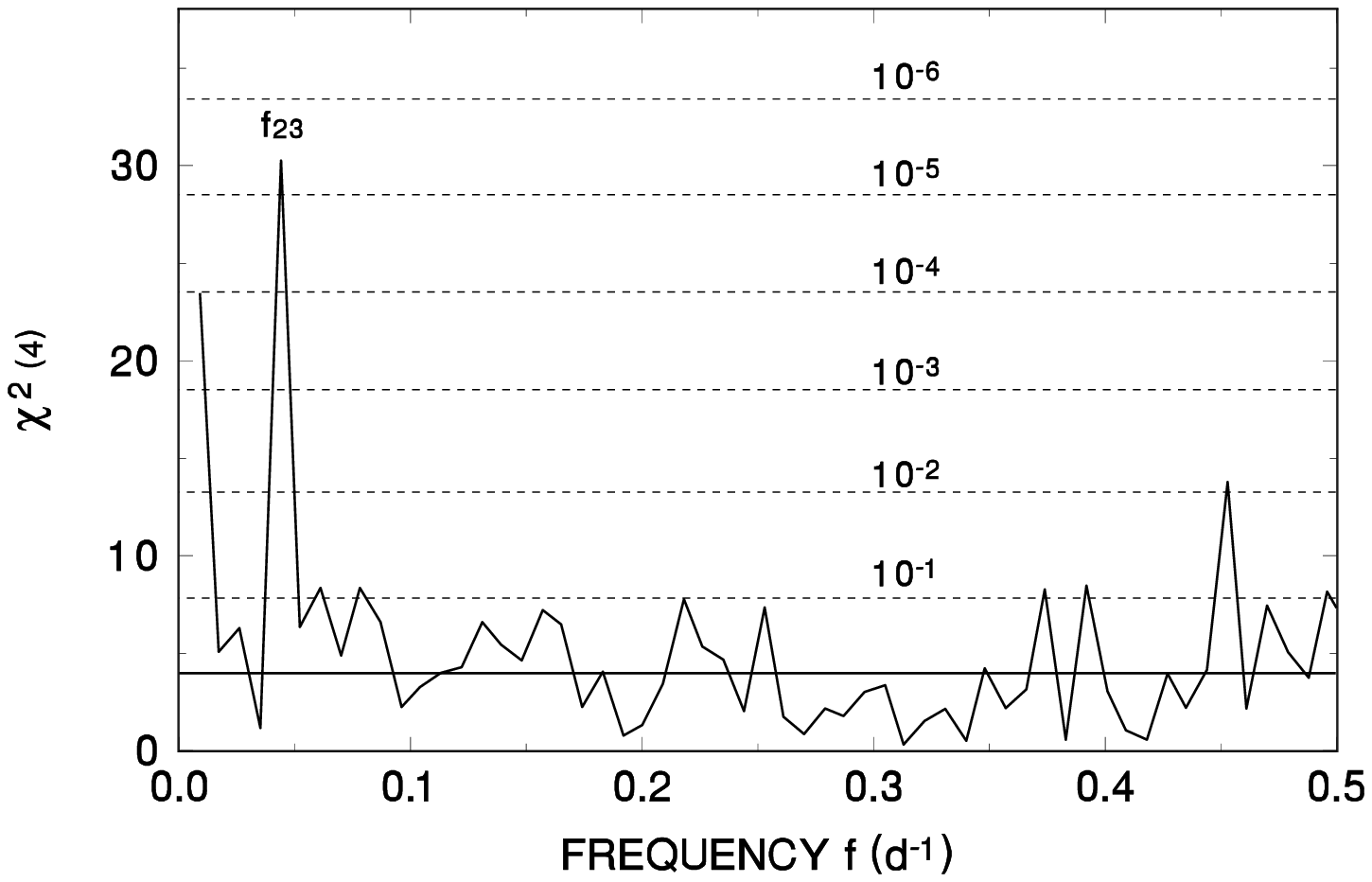,width=\linewidth,height=6.cm}
\caption{\label{comb_power_spectra}
The combined power spectrum (X-ray and TeV data) using proper normalization
(see text). The probability of the 23 day period, after multiplication
with the number of trials,  corresponds to 3.5 $\sigma$ .}
\end{minipage}
\end{figure}

The distribution of $P \left( \, \omega \, \right)$ is shown in
Fig.~\ref{norm_power_spectra}. The power values now follow the same $\exp
\, \left( \, -z \, \right)$ distribution as in the case of independent
Gaussian random noise (Equ.~\ref{prob_equ}). Again the 23 day period points
have big, but less significant deviations from this distribution. 

The final significance for the 23 day period is derived from the power
spectrum of the combined X-ray and TeV data. Here we use the fact that
if ${\cal P} \equiv \exp \, \left( \, -z \, \right)$, for each single
frequency
\begin{equation}
\chi^2\, \left( \,
4 \, \right) = -2 \, \ln \, \left( \, \exp \, \left(\, -z_\gamma \,
\right) \, \right) -2 \, \ln \, \left( \, \exp \, \left(\, -z_{\hbox{x-ray}} \,
\right) \, \right) = 2 \cdot (z_\gamma + z_{\hbox{x-ray}})
\end{equation}
is chi-square distributed with 4 degrees of freedom. The resulting
$\chi^2$ values together with the corresponding probability are
shown in Fig.~\ref{comb_power_spectra}. For the 23 day period a
probability ${\cal P} = 4.9\,e$--$\, 06$ corresponding to $4.4\, \sigma$ is
derived, which reduces to ${\cal P} = 2.8\, e$--$\, 04$ or $3.5\, \sigma$
after taking all 58 independent trial frequencies (between 1/T
and 0.5 per day) into account.

\section{Results and Conclusions}
The power spectrum shows a significant peak at 23 days in both the TeV
and X-ray data samples. This result is in accordance with the
significant correlation of 0.64 which has been observed between the
TeV and X-ray data (Aharonian 1999b). Using the framework of the shot
noise model the combined significance becomes 3.5 $\sigma$. In order
to avoid/reduce aliasing effects, introduced by the data gaps during
the moon period in the TeV range, moon observations have proven to be
of great importance.

\section*{Acknowledgments}
We acknowledge the rapid availability of the RXTE data. This work was
supported by the German Ministry of Education and Research, BMBF, the
Deutsche Forschungsgemeinschaft, DFG, and the Spanish Research Foundation,
CYCIT.

\vspace{1ex}
\begin{center}
{\Large\bf References}
\end{center}
%
Aharonian, F.A., et al., 1999a, A\&A 342, 69\\
Aharonian, F.A., et al., 1999b, A\&A accepted, see also astro-ph/9901248\\
Bradbury, S.M., et al., 1997, A\&A 320, L5\\
de Jager, O.C., Kranich, D., Lorenz, E. \& Kestel, M., 1999, these proceedings\\
Kranich, D., et al., 1999, Astroparticle Physics accepted, see also
astro-ph/9901330 \\
Lomb, N.R., 1976, Astrophysics and Space Science, 39, 447-462\\
Protheroe, R.J., et al., 1997, Proc. 25th ICRC, Durban, Vol. 8, 317\\
Quinn, J., et al., 1996, ApJ 456, L83 \\
Remillard, R.A. \& Levine, M.L., 1997, Proc. All Sky X-Ray Observations in
the Next Decade,\\
\indent see also astro-ph/9707338\\
RXTE, 1999, 'ASM/RXTE quick-look results', http://space.mit.edu/XTE/asmlc/ASM.html\\
Scargle, J.D., 1982, ApJ 263, 835\\

\end{document}